\documentclass[aps,pre,onecolumn,showkeys,superscriptaddress]{revtex4-2}

\usepackage{latexsym}
\usepackage{amsmath, amsthm, amssymb}
\usepackage{mathrsfs}
\usepackage{epsfig}
\usepackage{graphicx}
\usepackage{dcolumn}
\usepackage{expl3}
\usepackage{float}
\usepackage{xcolor}
\usepackage{natbib}
\usepackage{hyperref}
\usepackage{placeins}

\begin{document}

\title{Phototactic cyanobacteria as an active matter system}

\author{P. Varuni}
\thanks{These authors contributed equally to this work}
\affiliation{The Institute of Mathematical Sciences, CIT Campus, Taramani, Chennai 600113, India}
\author{Shakti N. Menon}
\thanks{These authors contributed equally to this work}
\affiliation{The Institute of Mathematical Sciences, CIT Campus, Taramani, Chennai 600113, India}
\author{Gautam I. Menon}
\thanks{corresponding author: gautam.menon@ashoka.edu.in}
\affiliation{The Institute of Mathematical Sciences, CIT Campus, Taramani, Chennai 600113, India}
\affiliation{Homi Bhabha National Institute, BARC Training School Complex, Anushaktinagar, Mumbai 400 094, India}
\affiliation{Departments of Physics and Biology, Ashoka University, Plot No. 2, Rajiv Gandhi Education City, NCR P.O.Rai, Sonepat Haryana 131 029, India}
\date{\today}

\begin{abstract}
Flocks of birds, schools of fish, mixtures of motors and cytoskeletal filaments, swimming bacteria and driven granular media are systems of interacting motile units that exhibit collective behaviour. These can all be described as active matter systems, since each individual unit takes energy from an internal energy depot and transduces it into work performed on the environment. We review a model for cyanobacterial phototaxis, emphasising the differences from other models for collective behaviour in active matter systems. The interactions between individual cells during phototaxis are dominated by mechanical forces mediated by their physical attachments through type IV pili (T4P) and through the production of ``slime'', a complex mixture of non-diffusible polysaccharides deposited by cells that acts to decrease friction locally. The slime, in particular, adds a component to the interaction that is local in space but non-local in time, perhaps most comparable to the pheromones laid down in ant trails. Our results suggest that the time-delayed component of the interactions between bacteria qualify their description as a novel active system, which we refer to as ``damp'' active matter.
\end{abstract}


\maketitle

\section{Introduction}

Organisms that aggregate in large groups often coordinate behaviour by sharing physical, chemical or spatial cues, culminating in collective emergent response \cite{Vicsek2012}. Such self-organised phenomena have been observed across a wide array of living systems, perhaps most spectacularly in flocks of birds \cite{Bialek2012,Cavagna2014}. Collective motion has also been observed and studied in other biological contexts including bacterial growth and taxis \cite{Krell2011}, cell migration during embryonic development \cite{Reig2014}, ant pheromone trails \cite{Jackson2006}, locust swarms \cite{Buhl2006}, fish schools \cite{Katz2011}, animal herds \cite{Couzin2005}  and even human crowds \cite{Helbing2000,Silverberg2013}. Individuals in such groups share information, likely benefiting from the effective distribution of available resources. 

In bacteria, collective motility has been observed under different conditions in many species \cite{Wolgemuth2002,Sokolov2007,Krell2011}. Examples of such motility  include phototaxis in cyanobacterial colonies. Cyanobacteria are photosynthetic microbes that process and respond to various wavelengths of light and exhibit phototaxis, or motion in response to light \cite{Schuergers2017}. The cyanobacterium \textit{Synechocystis} sp. PCC 6803 exhibits ``twitching'' or ``gliding'' motility, driven by Type IV pili (T4P). These long appendages facilitate physical connections between proximal cells. Apart from this, these cells extrude a mixture of complex polysaccharides, termed ``slime'', which reduces the friction \cite{Bhaya1999}. In this review, we highlight how cyanobacterial phototaxis can be viewed as a model for the movement of cell aggregates in response to complex stimuli.

A number of theoretical frameworks have been proposed to describe collective motility in living systems, including cellular automata, reaction-diffusion equations and agent-based models \cite{Vicsek2012,Mehes2014}. A versatile, and scale-independent, model for this type of self-organised motility is the flocking model proposed by Vicsek et al. \cite{Vicsek1995}. In its most general form, the model describes a group of agents (that could represent birds, fish, quadrupeds, bacteria, etc), whose direction of motion is governed by a set of internal rules that sense and respond to information related to their neighbourhood. This framework can also incorporate stochasticity in multiple ways, including by adding noise to information that is sent, received or processed by a single agent, as well as in the determination of the final direction of motion of each agent. 

These models belong to the larger class of active matter models, which have been used to describe the dynamics of a wide range of nonequilibrium systems \cite{Menon2010,Ramaswamy2010}. Such models describe the behaviour of systems of ``self-propelled particles'', i.e. which utilise an internal source of energy to move. These models are especially powerful in the context of living systems, since biological populations are inherently noisy and heterogeneous. Active matter models can provide a powerful tool to describe collective motility in such contexts. 

Active matter systems are often divided into two types \cite{Marchetti2013}: \textit{dry} (where agents only receive information from others in their neighbourhood) and \textit{wet} (where agents, in addition, take in local and global environmental cues). The main distinction between dry and wet models is that the latter incorporates hydrodynamic interactions mediated by the surrounding fluid medium, a biophysical effect of some importance in understanding phenomena like fish schooling and bacterial swimming. In cyanobacterial phototaxis, cells move over a surface and not in a fluid medium, thereby truncating the effects of the hydrodynamic interaction. However, through the accumulation of slime, the surface encodes a memory of cells that have traversed it, allowing for indirect cell-cell interactions that are retarded in time. We suggest that this time-delayed medium-dependent short-ranged interaction justifies a description in terms of a ``damp'' active matter system, distinct from both wet and dry active matter.

Collective behaviour in phototactic cyanobacteria presents a system where such active matter models can be tested and calibrated through relatively straightforward experimental observations and perturbations \cite{Menon2021}. These studies also provide insights into how single bacterial cells collect and integrate information to make decisions on an individual and group level. Models that incorporate noise and heterogeneity may better reflect bacterial colonies in the wild, providing insights into how natural populations coordinate behaviour. Since collective behaviour in this system relies on both physical interactions and local environment, it is markedly different from the other well-studied system that describes the coordination of population-level behaviour, namely quorum-sensing. Thus, cyanobacterial phototaxis presents consensus-building as a new paradigm for emergent collective behaviour in bacterial colonies.

\begin{figure}[t]
\centering
\includegraphics[width=\textwidth]{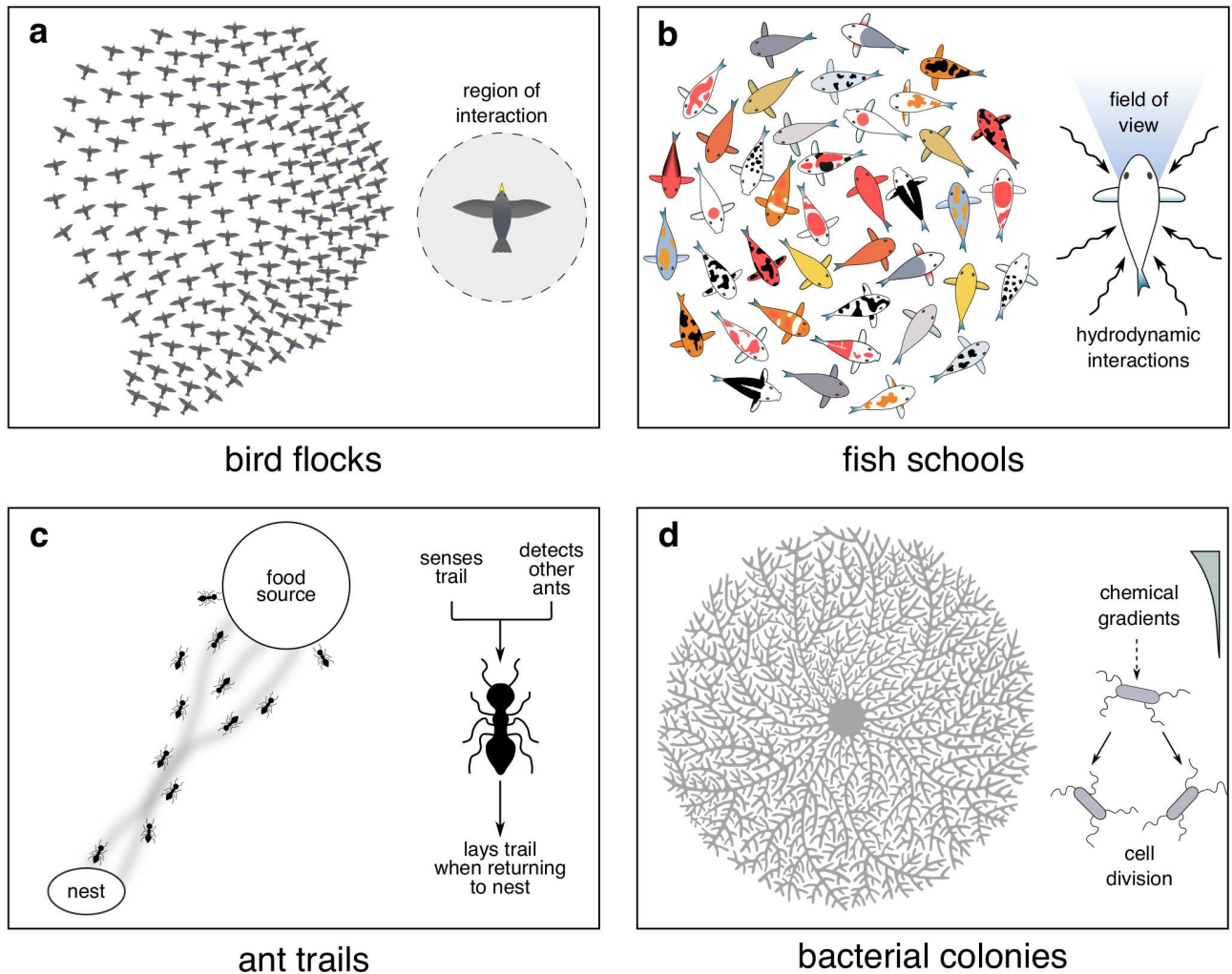}
\caption{
\textbf{Schematic illustrations of collective motion in a number of living systems.}
(a) A flock of birds: individual decisions are influenced by visual cues from other individuals in their local neighbourhood. (b) A school of fish: individual decisions are influenced by visual cues from other individuals in their local neighbourhood and long-range hydrodynamic interactions as result of motion of all other individuals in the medium. (c) Ant pheromone trails: individuals sense and follow pheromone trails laid down by preceding ants as well as follow other ants. (d) Bacterial colony growth: individuals divide and move to adjacent areas, depending on local individual density as well as local and global environmental cues.
}
\label{fig1}
\end{figure}

\section{Collective motion in living systems}

Bird flocking is one of the most widely-studied examples of collective motion in living systems. Although the emergent behaviour of this system is quite complex, most agent-based models that describe this phenomenon are relatively simple, where each agent makes decisions based strictly on the visual cues that it receives (Fig.~\ref{fig1}a). These decisions are typically based on the motion of other agents in the local neighbourhood. They are assumed not to be influenced by environmental factors. Furthermore, individual agents do not directly communicate with each other during flocking. Each agent thus processes information within its field of view, related to the position and/or direction of movement of neighbours, to make a decision on its own direction of movement. To capture features of the emergent behaviour observed in natural bird flocks, various models have also incorporated different aspects of the processing of visual information, including variability in the field of view \cite{Hemelrijk2012,Bagarti2019}, an explicit attraction to facilitate flock cohesion \cite{Gregoire2004}, and the role of social hierarchy within the flocks \cite{Nagy2010}.

Models of fish swarms typically incorporate a similar framework to bird flocking models, where the decisions regarding motion of each agent are informed by the motion of others within their field of view \cite{Lopez2012}. Since this collective motion occurs in a fluid, recent models have incorporated environmental cues that affect individual behaviour in the form of long range hydrodynamic interactions (e.g. \cite{Filella2018}) (Fig.~\ref{fig1}b). Here, the motion of each agent contributes to, and is affected by, the environmental flow field, wherein local perturbations propagate across the fluid. Thus, individual motion relies on not only the states of their neighbours but also on the dynamic environmental landscape that they contribute towards changing.

Ant pheromone trails are another interesting example of collective behaviour that have been modelled as individual agents that incorporate information from their local environment. Ants are known to deposit a chemical (pheromone) on the substrate as they move along, which then marks a trail for other ants to follow (Fig.~\ref{fig1}c). These pheromone signals can eventually dissipate, so the environmental ``memory'' is not permanent, although it can be reinforced as more and more ants traverse the trail that is already laid down. To model the effect of environmental memory, it is typically assumed that each agent decides on the direction and speed of motion, based on local environmental information that is encoded as memory in the state of the environment and is communicated across time \cite{Nishinari2006}. These trails have been found to allow ants to efficiently find and transport the food back to their nest \cite{John2009}.

Individual-based models have also been used to describe bacterial colony growth patterns, where cell decisions are not dependent on other individuals in their neighbourhood, but only on local environmental cues (Fig.~\ref{fig1}d). Here, cells are often modelled as ``walkers'' that can divide or move to new adjacent areas depending on local nutrients availability. In these models \cite{Benjacob1994,Benjacob1998,Benjacob2005}, the nature of the medium can also be incorporated through (1) the rate of diffusion of nutrients and (2) the ease at which cells can migrate to new regions, both which can be experimentally mirrored in agar concentration. These models show that changes in global parameters of the medium lead to different dynamics of local environmental landscapes,  finally resulting in emergent shapes and structures of colonies.

Cyanobacterial phototaxis is a collective phenomena where individuals can sense light \cite{Schuergers2016}, but their motion depends on physical interactions between cells and the local environment. Individual cells can attach to, and exert physical forces on, neighbouring cells through transient T4P attachments. Cells also secrete slime, which can change the local environmental landscape. In an active matter framework, these attachments are modelled as local physical interactions, similar to the field of view for birds or fish, while the slime is modelled as an environmental factor, similar to the hydrodynamic interactions in fish or local environmental variables in bacterial colony growth \cite{Varuni2017}. This model maintains local short-term interactions between individuals while also incorporating the effect of local long-term environmental memory on motion. 

Such models are qualitatively different from other models that have been studied as paradigms, the ``wet'' and ``dry'' active matter models. The time-retarded, slime-mediated interaction provides novel aspects to collective behavour in such systems, which we will refer to as ``damp'' active matter models.

\section{Modelling cyanobacterial phototaxis}

\begin{figure}[t]
\centering
\includegraphics[width=\textwidth]{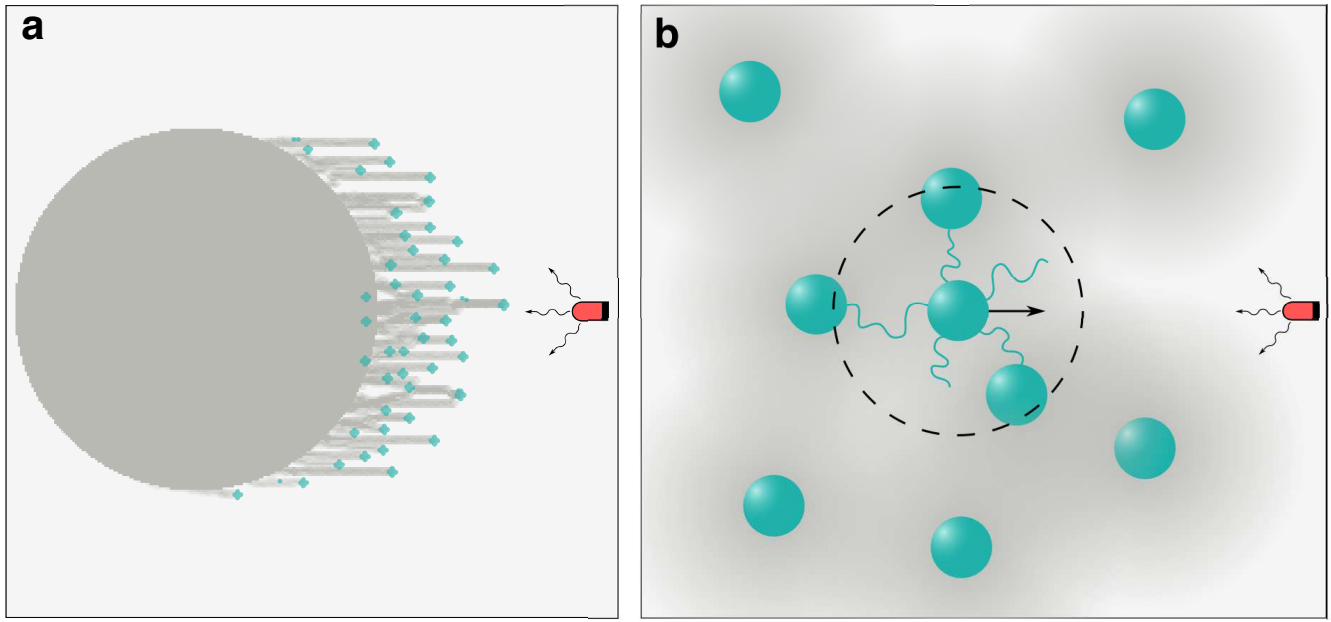}
\caption{
\textbf{Active matter model of cyanobacterial phototaxis.}
Emergent behaviour in a cyanobacterial colony, with cells indicated as green circles and underlying slime indicated in grey (darker grey implying more slime). (a) Cells are initially in a circular colony (large grey circle), which has maximal slime. On application of a directional light source (indicated as an LED, on the East), aggregates of cells emerge from the colony in finger-like projections. The figure is obtained by simulating the model in \cite{Varuni2017}. (b) (adapted from \cite{Varuni2017}) Schematic of individual cell this behaviour in this model: (i) cells displays a bias in motion towards the direction of the light source (ii) cells can exert forces on other cells that lie within a neighbourhood (dashed circle) (iii) cells lay down slime (iv) speed of the cell depends on the amount of slime at that position (more slime, faster the motion).
}
\label{fig2}
\end{figure}

\begin{figure}[t]
\centering
\includegraphics[width=\textwidth]{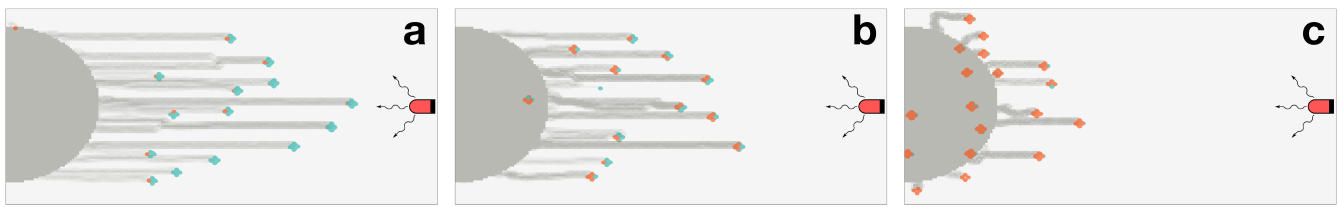}
\caption{
\textbf{Finger formation in colonies with ``freeloaders''.}
Simulations of populations with (a) $10\%$ (b) $50\%$ (c) $90\%$ of cells (indicated in red) cannot sense or respond to the light source (indicated as an LED).
Grey circular portion indicates initial colony with maximum slime.
}
\label{fig3}
\end{figure}

\begin{figure}[t]
\centering
\includegraphics[width=\textwidth]{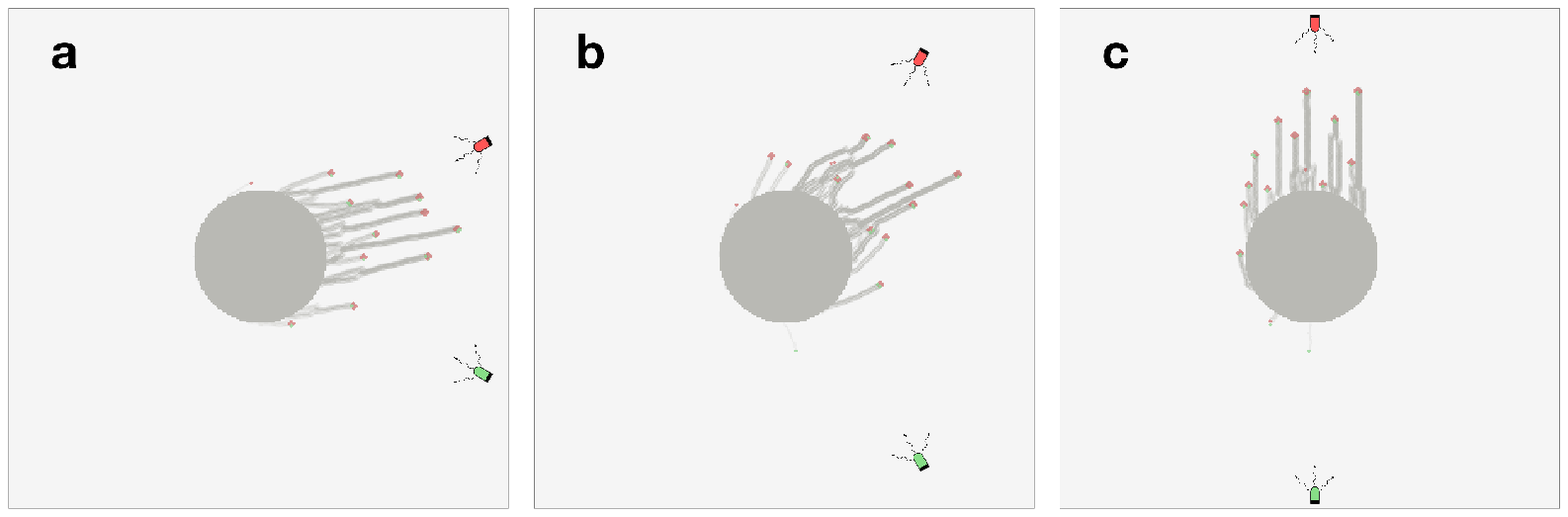}
\caption{
\textbf{Finger formation in heterogeneous colonies under different light scenarios}: Simulations of colonies comprising a mixture of two types of cells, $75\%$ of which only sense red light  (displayed in red) and the remaining $25\%$ only sense green light (displayed in green). Red and green lights (indicated as LEDs) are positioned at different angles North/South of East: (a) $30^\circ$ (b) $60^\circ$ (c) $90^\circ$.
Grey circular portion indicates initial colony with maximum slime.
}
\label{fig4}
\end{figure}

The phenomena of cyanobacterial phototaxis has been modelled using a number of approaches including cellular automata models, reaction-diffusion and active matter models. In one of the first agent-based models of cyanobacterial motility \cite{Galante2012} individual cells stochastically update their direction of motion based on the position of other cells in their vicinity, similar to the active matter framework of Vicsek et al. \cite{Vicsek1995}. Simulations of this model showed cell aggregations that recapitulate observations of the initial stages of phototaxis. Our recent model \cite{Varuni2017,Menon2020} builds on earlier frameworks to incorporate more detailed physical interactions between cells and environmental factors related to slime as well as external light stimuli. 


In this discrete-time model \cite{Varuni2017,Menon2020}, at every time step each cell independently makes a decision to move towards light with probability $p_{photo}$ and in a random direction with probability $1-p_{photo}$. Hence an isolated cell would effectively exhibit a biased random walk in the direction of light. However, in a colony setting, the motion of cells is modulated by physical interactions with neighbours through T4P, as well as the amount of underlying slime. To account for cell-cell T4P interactions, agents are allowed to transiently ``tug'' or apply an attractive force on neighbours that lie in their vicinity. Furthermore, to discourage overlaps between cells, the model incorporates soft-core interactions to implement a repulsive force. These cell-cell interactions are governed by a sigmoidal force term that is a function of the distance between the cell centres. To incorporate the effect of slime on motion, each agent adds to the local environment by ``depositing'' a non-diffusible resource, which reduces the friction of agents that subsequently traverse that region. To model the accumulation of slime, we consider a square lattice where at every time step each cell increments the slime content of the lattice site closest to its centre. Finally, the position of each is updated using the following expression:
\[\mathbf{X}_i^{t+1} = \mathbf{X}_i^{t} + \left(\mathbf{G}_i^{t} + \mathbf{F}_i^{t}\right)/\gamma_{(r,c)}^t\]
where $\mathbf{X}_i^{t}$ is the position of the cell in $\mathbb{R}^2$, $\mathbf{G}_i^{t}$ is the force associated with the decision of the cell to move in a particular direction, $\mathbf{F}_i^{t}$ is the total external force experienced by the cell and $\gamma_{c,r}^{t}$ is the friction that is modulated with the underlying slime at the lattice site $(c,r)$ that is closest to the centre of cell $i$. The versatility of this model is that it allows for different choices of parameter values that result in colony morphologies ranging from fast-moving finger-like projections to slow-moving fronts of cells.

Our model is able to capture the main features of experimentally observed phototaxis (Fig.~\ref{fig2}). At the initial stages of the simulation individual cells are relatively separated from each other and slime is in abundance in the colony. Due to their ability to sense light direction, these cells move more or less independently towards the colony edge closer to the light. As the cells begin to move they start to form small aggregates. When these aggregates collect at the edge of the colony, they merge to form dense fingers that extrude out of the colony towards the light source. This progression of colony-level phototaxis is in agreement with experimental observations \cite{Bhaya2006,Chau2015}. 

Our simulations indicate that even small biases in direction of motion, can lead to formation of robust fingers that move towards the light source. The distribution of direction of motion of individual cells indicate that even though individual cells may not move directly toward the light source at every step, the connections between cells and the narrow zone of motility limited by the presence of slime leads to motion that is biased more towards the light source than would be expected for isolated cells \cite{Varuni2017}. The resulting distribution of the direction of motion of individual cells in the fingers is similar to experimental observations \cite{Chau2015}. These results highlight how physical cell-cell interactions (through T4P) and a shared local resource (slime) can lead to a consensus in direction of motion.  


Active matter models allow for variation in behaviour of individuals in the group, and therefore can be conveniently used to investigate heterogeneous populations under complex stimuli. For example, as shown in Fig.~\ref{fig3}, we can investigate how a fraction of freeloaders can change characteristics of the speed and morphology of finger formation and progression \cite{Varuni2017}. Here, we define freeloaders to be cells that do not sense light, but can attach (or be attached) to other cells through T4P and can also lay down slime. Simulations indicate that such freeloaders are present in the aggregates and fingers that extend towards the light source, likely as a consequence of attachments to light-sensing cells which can tug them in the direction of light, and the lack of slime in other directions which restricts their movement outside the fingers. We see that even as the fraction of freeloaders increases, the finger formation and motion towards light remains relatively robust. Another example of the versatility of the model is that it allows us to study how colonies respond to complex illumination schemes \cite{Menon2020}. In particular, experimental observations of colonies under two simultaneous light signals indicate that cells can sense and integrate multiple signals \cite{Chau2017}. Investigating changes to the colony morphology and distribution of direction of motion in individual cells can also help us understand how individual cells integrate spatio-temporal information from multiple stimuli. In fact, such active matter models allow us to study the combination of heterogeneity and multiple stimuli by simulating colonies that have a mixture of cells, each sensitive to a particular wavelength of light (Fig.~\ref{fig4}). The flexibility of this modelling framework makes it suitable for investigating emergent cyanobacterial colony behaviour in experimental and natural settings.

\section{Outlook}

This review has summarised our recent work, placing it in a larger context that specifically highlights why cyanobacterial phototaxis should represent a novel system for active matter studies. Our model draws from biological evidence that suggests that cells can ``tug'' on their neighbours and ``deposit'' slime, which serves as a common resource that locally decreases friction \cite{Varuni2017,Menon2020}. The slime that is deposited has both the effect of making it easier for bacterial cells to move across surfaces as well as to allow these cells to interact across time, since bacteria can ride upon the slime earlier laid down by others. Therefore, in our model, direct interactions between agents are mechanical and instantaneous, while the indirect interactions between agents are local and time delayed. We hence suggest that this can be viewed as a ``damp'' active matter system, as it cannot be fully described by either the ``wet'' or ``dry'' frameworks. 

Our modelling framework allows us to investigate how individual cells respond to complex stimuli landscapes in the context of collective behaviour. This model of phototactic cyanobacteria recapitulates qualitative aspects of experimentally observed colony morphologies under different light wavelengths and intensities. We have also explored the extent to which individual information processing can affect emergent collective behaviour. When presented with multiple light sources, individual cells can either instantaneously sense and respond to a single light source or simultaneously sense and integrate their response to multiple light sources. Our results suggest that tracking trajectories of single cells, and not just colony morphology, should provide  a better understanding of decision-making at the single cell level \cite{Menon2020}. These insights could help parse how individual bacteria in natural environments can simultaneously sense and process multiple stimuli and then, integrate their response to them. Another major advantage of active matter models is that individuals can be assigned different decision-making capabilities. Indeed, we show how our model can easily incorporate heterogeneity, which could allow us to model wild populations. 

We suggest that in contrast to quorum-sensing, this ``damp'' active matter framework could be another paradigm of consensus-building that leads to collective behaviour in bacterial groups. These models could help us understand how bacterial populations in particular and living systems in general respond to complex stimuli through variable individual-level decision-making, which can then be combined and coordinated to yield collective behaviour. Establishing a stronger connection between experimental data and details of model construction should provide further insights into the emergence of collective behavior in a biological context.

\begin{acknowledgments}
PV is supported by Vigyan Pratibha (12th Plan), funded by the Department of Atomic Energy, Government of India. SNM is supported by the Centre of Excellence in Complex Systems and Data Science, funded by the Department of Atomic Energy, Government of India. 
\end{acknowledgments}



\end{document}